 \documentclass[aps,twocolumn,superscriptaddress,footinbib, reprint,longbibliography]{revtex4-2}
\usepackage{graphicx} 
\usepackage{url}
\usepackage{upgreek,amssymb,multirow}
\usepackage{graphicx,color}
\usepackage[dvipsnames]{xcolor}
\usepackage{mathtools}
\usepackage{float}
\usepackage{mciteplus}
\usepackage{blindtext}
\usepackage[unicode=true,pdfusetitle,bookmarks=false,linktocpage=true,colorlinks=true,citecolor=blue,urlcolor=blue,linkcolor=red]{hyperref}

\usepackage{amsmath}
\usepackage{extarrows}
\usepackage{braket}
\usepackage{tikz}
\usepackage{tikz-cd}

\newcommand{\bbra}[1]{\langle\!\langle #1|}
\newcommand{\kket}[1]{| #1 \rangle\!\rangle}

\newcommand{\Tr}{\operatorname{Tr}}

\begin{document}
\title{Gauge theory and mixed state criticality}

\author{Takamasa Ando}
\affiliation{Center for Gravitational Physics and Quantum Information, Yukawa Institute for Theoretical Physics, Kyoto University, Kyoto 606-8502, Japan}

\author{Shinsei Ryu}
\affiliation{Department of Physics, Princeton University, Princeton, New Jersey 08544, USA}

\author{Masataka Watanabe}
\affiliation{Graduate School of Science, The University of Tokyo, Tokyo 113-0033, Japan}

\begin{abstract}
In mixed quantum states, the notion of symmetry is divided into two types: strong and weak symmetry. While spontaneous symmetry breaking (SSB) for a weak symmetry is detected by two-point correlation functions, SSB for a strong symmetry is characterized by the R\'enyi-2 correlators. In this work, we present a way to construct various SSB phases for strong symmetries, starting from the ground state phase diagram of lattice gauge theory models. In addition to introducing a new type of mixed-state topological phases, we provide models of the criticalities between them, including those with gapless symmetry-protected topological order. We clarify that the ground states of lattice gauge theories are purified states of the corresponding mixed SSB states. Our construction can be applied to any finite gauge theory and offers a framework to study quantum operations between mixed quantum phases.
\end{abstract}

\maketitle

\tableofcontents

\section{Introduction}
Open quantum systems are quantum systems that interact with the environment, 
making them ubiquitous in nature.
While such interactions may be unwanted in certain practical applications, they can also be harnessed for beneficial outcomes when actively controlled. 
In order to do so, it is important to find unique phenomena that have no direct counterparts in closed systems.
For example, entangled states can be prepared by using dissipation and measurements \cite{2008NatPh...4..878D, Tantivasadakarn:2021vel, 
Verresen:2021wdv,Lu:2022jax,Lu:2023jmy,
Foss-Feig:2023uew,
Iqbal:2023shx,
Iqbal:2023wvm};
Topological phases and phenomena unique to open and non-hermitian quantum systems 
have also been classified  
\cite{Ashida:2020dkc,Okuma:2022bnb, RevModPhys.93.015005};
Quantum many-body systems under monitoring  
undergo a measurement-induced phase transition 
between distinct dynamical phases
\cite{Fisher:2022qey};
Furthermore, recent studies have explored 
the fate of topological phases
at finite temperature and under decoherence 
\cite{Dennis:2001nw,
Lee:2022hog,
Fan:2023rvp, 
Bao:2023zry,
Chen:2024knt},
as well as topological order intrinsic to mixed states 
\cite{Sohal:2024qvq, Ellison:2024svg, Wang:2023uoj, Kuno:2024bmf}.
With these exciting developments at hand, it would be important to have a unified method to understand and expand the landscape of those inherently open phenomena and phases of matter.

The notion of symmetry is crucial in understanding the behavior of many-body phases.
Indeed, symmetry has proven 
useful in classifying phases of matter of closed systems 
at equilibrium, such as spontaneous symmetry breaking (SSB) or 
symmetry-protected topological (SPT) phases
\cite{Gu:2009dr, Pollmann:2009mhk, Pollmann:2009ryx, Chen:2010zpc, Schuch:2010, Chen:2011pg, Levin:2012yb, Chen:2012ctz, Senthil:2014ooa}.
For open quantum systems, such a program is 
even more interesting 
due to the fact that symmetry can act from the left or the right 
on the mixed-state density matrix.
It is customary and convenient to classify the symmetry of the mixed state into two classes, strong and weak symmetries.
In the case of strong symmetry, the density matrix is (projectively) invariant under
the action of the left and the right symmetry independently, 
while in the weak symmetry case, 
only under the diagonal subgroup \cite{Albert_2014}.
Put differently, the former is defined by $U \rho = \rho U^{\dag}= \rho$, and the latter, less stringently, by $U \rho U^\dagger = \rho$, where $U$ is a unitary operator acting on the system’s Hilbert space, and $\rho$ the density matrix \footnote{All up to an overall phase factor accommodating the projective representation. Note also that we do not consider non-unitary symmetries in this paper.}.
We further define the spontaneous symmetry breaking of strong and weak symmetries by using off-diagonal long-range order, which will be discussed later in the main body of the text.
This distinction enables us to explore broader landscapes of quantum phases inherent to open quantum systems, such as SSB, phase transitions \cite{Lee:2023fsk,
Lessa:2024gcw, Sala:2024ply, Kuno:2024aiw, Shah:2024gzd} or SPT orders \cite{deGroot:2021vdi,Ma:2022pvq,  Ma:2023rji, Zhang:2022jul, Guo:2024iii, Ma:2024kma, Xue:2024bkt,You:2024mth,Sun:2024iwm}
Lieb-Schultz-Mattis type theorems
and quantum anomalies
for open quantum systems 
have also been formulated in
\cite{Kawabata:2023dlv, Wang:2024vjl, Hsin:2023jqa,Lessa:2024wcw}.

In this work, we present a unifying approach to classify phases of matter that are inherently open.
Concretely, we introduce a systematic 
way to construct 
various spontaneous symmetry-breaking phases of open quantum systems, 
including strong-to-weak SSB (SWSSB) phases,
by starting from 
the ground state phase diagram of
lattice gauge theory models which are closed.
Our construction also allows us to study 
criticalities between them,
including 
those that support gapless symmetry-protected topological (gSPT) order
\cite{Scaffidi:2017ppg, Verresen:2019igf, Thorngren:2020wet, Li:2022jbf, Li:2023knf, Yu:2021rng, Yu:2024toh, Zhang:2024doj, Wen:2022tkg, Wen:2023otf, Wen:2024qsg, Wen:2024udn, Huang:2023pyk, Huang:2024ror, Bhardwaj:2023bbf, Bhardwaj:2024qrf, Bhardwaj:2024ydc, Hidaka:2022bbz, Su:2024vrk, Myerson-Jain:2024, Ando:2024nlk}. 
By leveraging the structure of lattice gauge theories, we provide a framework for studying quantum operations between different mixed quantum phases, which provides 
valuable insights in 
open quantum many-body systems.

\section{$\mathbb{Z}_2$ gauge theory}
Let us start from lattice models with emergent gauge fields at low energy, such as the Hamiltonian lattice gauge theory 
\cite{Kogut:1979wt, Fradkin:1978dv}.
We in particular focus on the $\mathbb{Z}_2$ lattice gauge theory in one spatial dimension as a concrete example, whose Hamiltonian we denote by $H$.
Its degrees of freedom are composed of matter and gauge fields, which live on vertices (\it i.e., \rm sites) and links, respectively.
We denote the Hilbert space of the theory (before imposing the Gauss law constraint, to be introduced later) by
$\mathcal{H}\equiv 
\bigotimes_j \left(\mathcal{H}_{v,j} 
\otimes
\mathcal{H}_{l,j+1/2}\right)
$,
where ${\cal H}_{v,j}$ is the matter Hilbert space at site $j$, while ${\cal H}_{l,j+1/2}$ is the 
$\mathbb{Z}_2$ gauge field Hilbert space living on the link connecting sites $j$ and $j+1$, both of which are two-dimensional.
We denote the Pauli operators on ${\cal H}_{v,j}$ as $X_j$, $Y_j$ and $Z_j$, and on ${\cal H}_{l,j}$ as $\sigma_j^x$, $\sigma_j^y$ and $\sigma_j^z$.

Physical states in the $\mathbb{Z}_2$ gauge theory, as in any gauge theories, 
satisfy the Gauss law constraint.
This is a constraint that $\sigma^z_{j-1/2}X_{j} \sigma^z_{j+1/2}=1$ for any physical states $\ket{\psi}$, \it i.e., \rm
\begin{equation} \label{gauss}
    G_j\ket{\psi}=+\ket{\psi},
    \quad 
    G_j=
    \sigma^z_{j-1/2}X_{j} \sigma^z_{j+1/2},
\end{equation}
for all $j$.
This can indeed be interpreted as gauging the $\mathbb{Z}_2$ symmetry of the matter theory --
The would-be $\mathbb{Z}_2$ global symmetry of the matter theory generated by $\prod_j X_j$ (which flips the spin at the vertices all at once) acts on physical states trivially as $\prod_j X_j=\prod_j G_j$.
(Throughout the paper, 
we will work with periodic boundary conditions unless stated 
otherwise.)

One can also impose such a constraint energetically in the UV lattice model, by explicitly adding $-K \sum_j G_j$ to the Hamiltonian -- If $K$ is large enough, we get the same ground state as the original $\mathbb{Z}_2$ gauge theory.
In the following, we will call this procedure the effective gauging and the resulting theory as the effective gauge theory \cite{Borla:2020avq, Verresen:2022mcr, Ando:2024nlk, com:effective_gauge, Xu:2023zsz}.
As it has been proven useful in constructing lattice models with SPT orders 
\cite{Gu:2009dr, Pollmann:2009mhk, Pollmann:2009ryx, Chen:2010zpc, Schuch:2010, Chen:2011pg, Levin:2012yb, Chen:2012ctz, Senthil:2014ooa}
or gapless topological phases 
\cite{Scaffidi:2017ppg, Verresen:2019igf, Thorngren:2020wet, Li:2022jbf, Li:2023knf, Yu:2021rng, Yu:2024toh, Zhang:2024doj, Wen:2022tkg, Wen:2023otf, Wen:2024qsg, Wen:2024udn, Huang:2023pyk, Huang:2024ror, Bhardwaj:2023bbf, Bhardwaj:2024qrf, Bhardwaj:2024ydc, Hidaka:2022bbz, Su:2024vrk, Myerson-Jain:2024, Ando:2024nlk} in closed equilibrium systems,
we will utilize it to study analogous phases in open quantum systems as well.
It also has an advantage over the usual lattice gauge models as there is no need to impose Gauss law and hence all the symmetries are global symmetries.

\section{Main claim}
We are interested in various phases of matter realized in mixed states, obtained by tracing out the matter degrees of freedom in the ground state pure state of lattice gauge theories, since they already exhibit various interesting phases at equilibrium and we expect this to carry over to mixed states.
We denote the environment Hilbert space by $\mathcal{H}_A$, which is a subspace of $\mathcal{H}_v$ on which the symmetry acts faithfully.
In our examples below, 
$\mathcal{H}_A=\mathcal{H}_v$
in the first example, 
while in the second and third, $\mathcal{H}_A 
\subsetneq \mathcal{H}_v$.

Our claim is that, the operation of taking the partial trace, which is commonly used in studying the phases of the mixed state, can be replaced by quantum channels describing decoherence and gauge fixing (or \it vice versa\rm ):

For a density matrix $\rho$ composed of pure states satisfying the Gauss law, \it i.e., \rm
$G_j \rho = \rho G^{\dag}_j =\rho$,
we have
\begin{equation}\label{eq:main_eq}
\varrho:=
    \operatorname{Tr}_{\mathcal{H}_{A}}(\rho)=\mathcal{E}_{ZZ}\left(\operatorname{Tr}_{\mathcal{H}_{A}}(\mathcal{E}_{CZ}(\rho))\right).
\end{equation}
Here, $\mathcal{E}_{ZZ}$ and $\mathcal{E}_{CZ}$ are quantum operations acting on density matrices defined in the following way, 
\begin{gather}
    \mathcal{E}_{CZ}(\rho)\coloneqq U_{CZ}\rho \,U_{CZ}^\dagger,\\
    U_{CZ}=\prod_{j=1}^{L}\texttt{CZ}_{j-1/2,j}\texttt{CZ}_{j,j+1/2},
\end{gather}
and 
\begin{gather}
    \mathcal{E}_{ZZ}(\rho)\coloneqq \left(\cdots\mathcal{E}_{ZZ,j}\circ \mathcal{E}_{ZZ,j+1}\circ\cdots\right)(\rho),\\ \mathcal{E}_{ZZ,j}(\rho)=\frac{\rho+
    \sigma^z_{j-\frac{1}{2}} \sigma^z_{j+\frac{1}{2}}\rho\,
    \sigma^z_{j-\frac{1}{2}} \sigma^z_{j+\frac{1}{2}}}{2},
\end{gather}
where $\texttt{CZ}_{j,k}$ is the contorolled-Z gate between two qubits. 
More concretely, we define $\texttt{CZ}_{j,k}$ as
$
    \texttt{CZ}_{j,k}=\mathop{\mathrm{diag}}(1,1,1,-1) 
$
on the $(\ket{\uparrow_j,
\uparrow_{k}},
\ket{\uparrow_j,
\downarrow_{k}},
\ket{\downarrow_j,\uparrow_k
},
\ket{\downarrow_j,
\downarrow_k})$ basis system.

Deferring the proof to Appendix \ref{sec:proof_main}, let us now discuss the physical meaning of \eqref{eq:main_eq}.
First of all, what the operation $\rho\mapsto \Tr_{\mathcal{H}_A}(\mathcal{E}_{CZ}(\rho))$ achieves is the gauge fixing on each pure state included in $\rho$.
This is because $\mathcal{E}_{CZ}$ transforms the Gauss constraint 
$\sigma^z_{j-{1/2}}X_{j}\sigma^z_{j+1/2}=1$ to $X_{j}=1$, and hence the matter degrees of freedom are disentangled from the rest. 
We call this gauge the unitary gauge. 
(See Appendix \ref{sec:review_gauge} for more discussions.)
Precisely speaking, the exact disentangling only happens for lattice gauge theories in which the gauge symmetry is manifest in the UV -- for effective gauge theories, there would be a term to freeze the matter spin $X_{j}$ to $1$.
When such a term becomes larger and larger in the IR, there is an effective disentangling between the matter and the gauge degrees of freedom.

\begin{table*}[htbp]
    \centering
    \caption{Each phase 
    in the unitary gauge.}
    \begin{tabular}{|c|c|c|c|} \hline
        model & $J<1$ & $J=1$ & $J>1$\\ \hline
        $H_1$ & $\mathbb{Z}_2$ SSB & Ising CFT & $\mathbb{Z}_2$ trivial\\ 
        $H_2$ & $\mathbb{Z}_2$ trivial $\times~\mathbb{Z}_2$ SSB & gSPT & $\mathbb{Z}_2\times\mathbb{Z}_2$ SPT\\ 
        $H_3$ & $\mathbb{Z}_4$ trivial $\times~\mathbb{Z}_2$ SSB & igSPT & $(\mathbb{Z}_4\rightarrow\mathbb{Z}_2^A$ SSB) $\times~(\mathbb{Z}_2^A\times\mathbb{Z}_2^B$ SPT)\\ \hline
    \end{tabular}
    \label{tb:pre_dph}
\end{table*}
\begin{table*}[htbp]
    \centering
    \caption{Mixed phases after implementing the operation $\mathcal{E}_{ZZ}.$}
    \begin{tabular}{|c|c|c|} \hline
        model & $J<1$ & $J>1$\\ \hline
        $H_1$ & $\mathbb{Z}_2$ SSB  & $\mathbb{Z}_2$ SWSSB\\ 
        $H_2$ & $\mathbb{Z}_2$ trivial $\times~\mathbb{Z}_2$ SSB & SWSSB-ASPT\\ 
        $H_3$ & $\mathbb{Z}_4$ trivial $\times~\mathbb{Z}_2$ SSB & $(\mathbb{Z}_4\rightarrow\mathbb{Z}_2$ SSB) $\times$ (SWSSB-ASPT)\\ \hline
    \end{tabular}
    \label{tb:dphd_phase}
\end{table*}

\begin{table*}[htbp]
    \centering
    \caption{Nontrivial order parameters at each criticality. $\braket{\cdot}_{4}$ denotes four-point correlators (R\'enyi-2 correlators). $\Delta$ and $\Delta_3$ are scaling dimensions of the corresponding operators of the criticalities.}
    \begin{tabular}{|c|c|} \hline
    $1$ & 
    $\braket{\sigma^z_{i}\sigma^z_{i+r}}=O\left(1/|r|^{2\Delta}\right), \quad 
    \braket{\sigma^z_{i}\sigma^z_{i+r}}_{4}=1$
    \\
    $2$ & 
    $\braket{\sigma^z_{i}\sigma^z_{i+r}}
    =O\left(1/|r|^{2\Delta}\right),
    \quad 
    \braket{\sigma^z_{i}\sigma^z_{i+r}}_{4}=1,
    \quad 
    \braket{\sigma^z_{i-1/2}X_{i}
    \, \cdots\, 
    X_{i+r} \sigma^z_{i+r+1/2}}=O(1)$\\ $3$ & $\braket{\tau_{i}^z\tau_{i+r}^z}=O\left(1/|r|^{2\Delta_3}\right), \quad \braket{\tau_{i}^z\tau_{i+r}^z}_{4}=1,\quad \braket{\tau_{i-1/2}^z \hat{X}_{i}^2\cdots \hat{X}_{i+r}^2 \tau_{i+r+1/2}^z}=O(1)$\\ \hline 
    \end{tabular}
    \label{tb:correlation}
\end{table*}

\begin{table*}[htbp]
    \centering
    \caption{The correspondence between
    the density matrix and doubled state pictures.
    \label{Table IV}
    }
    \begin{tabular}{|c|c|c|c|}\hline
        Density matrix 
        & 
        2-point $\operatorname{Tr}(\rho\,\sigma_{i}^z\sigma_{i+r}^z)$ 
        & 4-point $\operatorname{Tr}(\rho\,\sigma_{i}^z\sigma_{i+r}^z\rho\,\sigma_{i}^z\sigma_{i+r}^z)$ 
        & $\operatorname{Tr}(\rho^2 \sigma^z_{i} \sigma^z_{i+r})$\\
        \hline
        Doubled state 
        & strange correlator 
        & correlator for off-diagonal symmetry 
        & 
       correlator 
        \\ 
      &
     $\bbra{\rho_0}\sigma^z_{i}\sigma^z_{i+r}\kket{\rho}$ 
      &
       $\bbra{\rho}\sigma_{i}^z\tilde{\sigma}_{i}^z\sigma_{i+r}^z\tilde{\sigma}_{i+r}^z \kket{\rho}$ 
      &
      $\bbra{\rho}\sigma_{i}^z\sigma_{i+r}^z\kket{\rho}$
      \\
        \hline
    \end{tabular}
\end{table*}

Our main result \eqref{eq:main_eq} can be useful in different ways.
First of all, we can use it to study the effect of decoherence on $
\tilde{\rho}=
\mathrm{Tr}_{{\cal H}_A}\, \left(
{\cal E}_{CZ}\,(\rho)
\right)
$. (Note that $\tilde{\rho}$ is a pure state from the discussion above.)
This is useful because $\tilde{\rho}$ can be thought of as the ground state of the new (non-gauge) Hamiltonian $\tilde{H}$, which can be obtained from $H$ algorithmically by gauge fixing and a projection.
We will discuss this shortly using examples in the next section.

The claim \eqref{eq:main_eq} can also be used to infer properties of the mixed state (reduced) density matrix $\varrho$.
For example, it is immediate that $\varrho$ spontaneously breaks the strong $\mathbb{Z}_2$ symmetry of flipping the gauge spin.
The spontaneous symmetry breaking in this paper will be defined by using the so-called R\'enyi-2 correlator \cite{Lee:2023fsk, Lessa:2024gcw,Sala:2024ply, 
Weinstein:2024fug},
such that the spontaneous symmetry breaking happens when
\begin{equation}
    \lim_{r\to\infty}\frac{\operatorname{Tr}\left(\varrho \sigma^z_{i}\sigma^z_{i+r}\varrho 
    \sigma^z_{i}\sigma^z_{i+r}\right)}{\operatorname{Tr}\left(\varrho^2\right)}\neq 0.
\end{equation}
To see this, first notice that $\mathcal{E}_{ZZ}(\tilde{\rho})$ is symmetric under the strong $\mathbb{Z}_2$ symmetry as long as $\tilde{\rho}$ is symmetric as well, which is the case for us.
Moreover, one can see that two-point correlations are exactly preserved under the operation, 
\begin{equation}
\braket{\sigma^z_{i}\sigma^z_{i+r}}\coloneqq\operatorname{Tr}\left(
\varrho
\sigma^z_{i}\sigma^z_{i+r}\right)
    \propto\operatorname{Tr}\left(\tilde{\rho} \sigma^z_{i}\sigma^z_{i+r}\right)
\end{equation}
as 
\begin{equation}
    \mathcal{E}_{ZZ}(\rho)\propto\sum_{\{s\}_{j}}
    \prod_{j\in \mathbb{Z}+1/2}
    (\sigma^z_{j})^{s_{j}}\rho
    \prod_{j\in \mathbb{Z}+1/2}
    (\sigma^z_{j})^{s_{j}}. 
\end{equation}
Here, $s_{j}\in\{0,1\}$ and the summand for $\{s\}_{j}$ runs over all configurations such that the number of $j$ with $s_{j}=1$ is even.
We also see that the R\'enyi-2 correlations are always nontrivial,
\begin{equation}
    \frac{\operatorname{Tr}\left(\varrho \sigma^z_{i}\sigma^z_{i+r}\varrho 
    \sigma^z_{i}\sigma^z_{i+r}\right)}{\operatorname{Tr}\left(\varrho^2\right)}=1,
\end{equation}
by using the form of $\mathcal{E}_{ZZ}(\rho)$ -- The density matrix $\varrho$ breaks the strong $\mathbb{Z}_2$ symmetry spontaneously as promised.

Strikingly, if the procedure is applied to gauge theories at criticality, we always end up with some sort of critical points for open systems.
In the following, we explore three classes of models with such criticalities.

\section{Criticality between SWSSB and SSB} 
We first explore the simplest criticality that is intrinsic to mixed states; between SWSSB and SSB phases. As a prime example, we investigate the criticality of the transverse-field Ising (TFI) model. 
The Hamiltonian for the gauge theory is given by 
\begin{align}\label{eq:TFI_eff}
\begin{split}
    H_{1}\coloneqq 
    &-\sum_{j=1}^{L}\left(X_{j}+J Z_{j} \sigma^x_{j+1/2}Z_{j+1} \right)\\
    &\quad\quad\quad -K\sum_{j=1}^{L}
    \sigma^z_{j-1/2}X_{j} \sigma^z_{j+1/2},
    \end{split}
\end{align}
where
$L$ is the total number of lattice sites and we impose the periodic boundary condition,
and we take $J>0$ 
throughout the discussion.
In this expression, the model has 
two $\mathbb{Z}_2$ global symmetries generated by $\prod_{j}\sigma^x_{j+1/2}$ and $\prod_{j}X_{j}$. 
Depending on the value of the parameter $J$ the ground state of \eqref{eq:TFI_eff} belongs to three different phases (Table \ref{tb:pre_dph}). 
When $J<1$, 
the first $\mathbb{Z}_2$ symmetry 
is spontaneously broken in the ground state, while the other remains unbroken. 
When $J>1$, the ground state exhibits 
the $\mathbb{Z}_2\times\mathbb{Z}_2$ SPT order. 
While the model is gapless at $J=1$, 
it also has a nontrivial string order correlation 
$\braket{\sigma^z_{i-1/2}X_{i}\, \cdots\, X_{i+r} \sigma^z_{i+r+1/2}}$. 
Moreover, this model has a protected edge mode when put on the open boundary.
Gapless systems 
that exhibit these features are 
known as gapless SPT (gSPT) phases \cite{Scaffidi:2017ppg, Verresen:2019igf}.

We now discuss various mixed states obtained from the ground state by tracing the vertex degrees of freedom
(Table \ref{tb:dphd_phase}).
When tracing out the vertex degrees of freedom, the SSB (SPT) phase becomes the SSB (SWSSB) phase 
\cite{Sala:2024ply}.
Therefore, we obtain a critical mixed state between SSB and SWSSB phases,
by tracing out the gapped degree of the gSPT state at $J=1$.

Let $\rho_0$ be a ground state of $H_1$. Then $\widetilde{\rho_0}=\operatorname{Tr}_{\mathcal{H}_{A}}\left(\mathcal{E}_{CZ}(\rho_0)\right)$ is a ground state of the gauge theory 
in the unitary gauge. Specifically, 
$\widetilde{\rho_0}$ is a ground state of the Hamiltonian 
\begin{equation}
\label{model 1 unitary gauge}
    \widetilde{H_{1}}
    =-\sum_{j=1}^{L}
    \left(
    \sigma^z_{j-1/2}\sigma^z_{j+1/2}
    +J \sigma^x_{j+1/2}\right),
\end{equation}
which exhibits
the unbroken, gapped phase ($J>1$),
the SSB phase ($J<1$),
and the Ising criticality separating them ($J=1$).
%
The two-point correlation function of the ground state behaves as 
\begin{equation}
    \operatorname{Tr}(\widetilde{\rho_0} \sigma^z_{i}\sigma^z_{i+r})\xrightarrow{r\rightarrow\infty}
    \begin{cases}
        O(1),\quad &J<1,\\
        e^{-m|r|}, \quad &J>1,\\
        O\left(\frac{1}{|r|^{2\Delta}}\right), \quad &J=1.
    \end{cases}
\end{equation}
in the thermodynamic limit, where $\Delta$ is the scaling dimension of $\sigma^z$ at criticality, while at $J>1$, $m$ is the gap of the system. This correlation 
remains the same after the $\mathcal{E}_{ZZ}$ operation.
On the other hand, as discussed, 
the R\'enyi-2 correlators are nontrivial for any $J$.
Namely, the $J>1$ phase is mapped to the SWSSB phase under the $\mathcal{E}_{ZZ}$ operation. As for the critical point $J=1$, $\operatorname{Tr}_{\mathcal{H}_A}(\rho_0)=\mathcal{E}_{ZZ}(\widetilde{\rho_0})$ exhibits the criticality between SWSSB and SSB phases. %
We summarize various correlation functions of this model at criticality in Table \ref{tb:correlation}.

\section{Criticality between ``SWSSB-ASPT'' and SSB}
In the previous example, we discussed the model that exhibits Ising CFT criticality in the corresponding gauge theory, and such criticality describes the transition between the $\mathbb{Z}_2$ SSB phase and the $\mathbb{Z}_2$ trivial phase. 
Let us now discuss a criticality between an SSB phase and a nontrivial SPT phase. Such a criticality is described by a gSPT. 
As a model for $\mathbb{Z}_2$ (effective) gauge theory, we consider the following Hamiltonian:
\begin{align}
\label{H2}
        H_{2}=-\sum_{j=1}^{L}&
        \left(X_{j}
        +J\tau_{j}^z
        Z_{j} 
        \sigma^x_{j+1/2}\tau_{j+1}^z
        Z_{j+1}
        +K_0\tau_{j}^x X_{j}\right)
        \nonumber\\
        &-K\,\sum_{j=1}^{L}
        \sigma^z_{j-1/2}
        \tau_{j}^x 
        \sigma^z_{j+1/2},
\end{align}
where $\tau_{j}^{x,y,z}$ are 
the Pauli matrices,
$K_0$ is a sufficiently large positive constant,
and the last term (positive $K$) is for effective gauging.
We note that 
$\{\tau_j\}_j$ represents the matter degrees of freedom, 
which can be removed in the unitary gauge.
In the unitary gauge
or after implementing $\operatorname{Tr}_{\mathcal{H}_A}\left(\mathcal{E}_{CZ}(\cdot)\right)$ operation, the ground state of the model is the same as the following Hamiltonian:
\begin{equation}\label{eq:gSPT_pure}
    \widetilde{H_{2}}
    =-\sum_{j=1}^{L}
    \left(X_{j}
    +J Z_{j} \sigma^x_{j+1/2}
    Z_{j+1}
    +K_0\,\sigma^z_{j-1/2}
    X_{j} \sigma^z_{j+1/2}\right).
\end{equation}
This model is the same as \eqref{eq:TFI_eff}, 
and the critical point $J=1$ 
separates the SSB and SPT phases 
(Table \ref{tb:pre_dph}).
Since the last term commutes with the other terms, 
this term is stabilized at the ground state 
and gives a gapped sector. 
Looking at the gapless low-energy sector, this model is described by the Ising CFT. 
There are two correlators that characterize this critical point. 
One of them is 
$\braket{\sigma^z_{i}\sigma^z_{i+r}}$ and it exhibits algebraic decay which corresponds to the two-point correlation in the Ising CFT, 
and the other is the string correlator $\braket{\sigma^z_{i-1/2}X_{i}\, \cdots\, X_{i+r} \sigma^z_{i+r+1/2}}$ that indicates long-range order. 

Since operators that consist of the quantum operation $\mathcal{E}_{ZZ}$ commute with the two correlators, these correlations are preserved under the operation.
On the other hand, 
the quantum operation can affect the R\'enyi-2 correlation for $\sigma^z_{i}\sigma^z_{i+r}$, 
and indeed this exhibits long-range order after the operation. 
We realize that,
after taking the partial trace of the ground state 
of \eqref{H2},
$
\mathrm{Tr}_{{\cal H}_A}\, 
(\rho)$,
we have an interesting 
mixed state
for $J>1$.
While 
$\mathrm{Tr}_{{\cal H}_A}\, 
(\rho)$
exhibits 
a strong-to-weak SSB order
with respect to 
$\prod_j \sigma^x_{j+1/2}$, there still be a non-vanishing string correlation $\braket{\sigma^z_{i-1/2} X_{i}\, 
\cdots\, X_{i+r} \sigma^z_{i+r+1/2}}$. 
We note that the other string correlator $\braket{Z_{i} \sigma^x_{i+1/2}\, 
\cdots\, \sigma^x_{i+r-1/2} Z_{i+r}}$, which also characterizes the nontrivial SPT phase, no longer exhibits long-range order after the $\mathcal{E}_{ZZ}$ operation. 
Such a behavior of the two string correlations 
is one of the hallmarks of average SPT (ASPT) phases \cite{Ma:2022pvq, Ma:2023rji}. 
In this sense, the phase for $J>1$ is kind of a ``mixture'' of SWSSB and ASPT phases. We call this phase \textit{SWSSB-ASPT} phase.
In summary, the two gapped phases of the ground state of \eqref{eq:gSPT_pure} are mapped to the SSB ($J<1$) and SWSSB-ASPT ($J>1$) phases under the quantum operation $\mathcal{E}_{ZZ}$, 
and at $J=1$ the critical point 
is mapped 
to the critical mixed state 
between them
(Table \ref{tb:dphd_phase}).
We summarize various correlation functions of this model at criticality in Table \ref{tb:correlation}.

\section{Criticality between different (SW)SSB patterns}
Let us explore a critical model obtained by applying the $\mathcal{E}_{ZZ}$ operation to 
an intrinsically gapless SPT (igSPT) criticality \cite{Thorngren:2020wet}. Since igSPT models exhibit emergent 't Hooft anomalies in the IR and such anomalies forbid the existence of unique gapped ground states, such criticality may describe phase transitions between different SSB patterns.
Let us consider the model of the effective gauge theory of the form 
\begin{align}
        H_{3}=-\sum_{j=1}^{L}&\left(\hat{X}_{j}+J\hat{Z}_{j}\tau_{j}^z\sigma_{j+1/2}^x\tau_{j+1}^z\hat{Z}_{j+1}^\dagger+K_0\hat{X}_{j}^2\tau_{j}^x\right)
        \nonumber \\
        &-K\sum_{j=1}^{L}\sigma_{j-1/2}^z\tau_{j}^x\sigma_{j+1/2}^z+\mathrm{h.c.}
\end{align}
Here, $K_0$ is again a sufficiently large positive constant and $\hat{X}_{j},\hat{Z}_{j}$ are generalized Pauli matrices acting on a $j$-th four-dimensional qudit space and satisfy
$\hat{X}_{j}^4=\hat{Z}_{j}^4=1,\quad \hat{Z}_{j}\hat{X}_{j}=i\hat{X}_{j}\hat{Z}_{j}.
$
In the unitary gauge, this model is written as
\begin{align}\label{eq:igSPT_pure}
    \begin{split}
        \widetilde{H_{3}}=-\sum_{j=1}^{L}&\left(\hat{X}_{j}+J\hat{Z}_{j}\sigma_{j+1/2}^x \hat{Z}_{j+1/2}^\dagger\right.\\
        &\left.+K_0\,\sigma_{j-1/2}^z \hat{X}_{j}^2\sigma_{j+1/2}^z\right)+\mathrm{h.c.}
    \end{split}
\end{align}
At $J=1$, 
this model is known to realize an igSPT phase with a $\mathbb{Z}_4\times\mathbb{Z}_2$ global symmetry \cite{Su:2024vrk,Ando:2024nlk}. The $\mathbb{Z}_4$ symmetry is generated by $\prod_{j}\hat{X}_{j}$,
while the $\mathbb{Z}_2$ symmetry is generated by $\prod_{j}\sigma_{j+1/2}^x$. 
When $J<1$, 
the $\mathbb{Z}_4$ symmetry remains  unbroken,
while the $\mathbb{Z}_2$ symmetry is spontaneously broken. 
When $J>1$, the $\mathbb{Z}_4$ symmetry breaks to the $\mathbb{Z}_2$ subgroup while the other $\mathbb{Z}_2$ global symmetry remains unbroken. Moreover, a nontrivial SPT phase with respect to this unbroken $\mathbb{Z}_2\times\mathbb{Z}_2$ symmetry is stacked.
At $J=1$, the low-energy sector of the model is described by the $U(1)_4$ CFT \footnote{The $U(1)_4$ CFT is Tomonaga-Luttinger liquid with a certain Luttinger parameter.}. 
See Table \ref{tb:pre_dph} for the phase diagram.
The criticality of the corresponding CFT is captured by two-point correlations e.g.~$\braket{\sigma_{i}^z\sigma_{i+r}^z}$. In addition to such usual CFT correlators, this igSPT is characterized by the string order parameter $\braket{\sigma_{i-1/2}^z\hat{X}_{i}^2\, \cdots\, \hat{X}_{i+r}^2\sigma_{i+r+1/2}^z}$. One can see this correlation has an $O(1)$ expectation value because the last term in \eqref{eq:igSPT_pure} commutes with the other term, and so it is stabilized at the ground state.

Let us consider the mixed state obtained by applying the operation 
$\mathcal{E}_{ZZ}$
(Table \ref{tb:dphd_phase}). 
For $J<1$ where the $\mathbb{Z}_2$ symmetry is spontaneously broken, the state is mapped to the same SSB phase.
In contrast, for $J>1$, the state is mapped to an SSB phase where the  $\mathbb{Z}_4$ symmetry breaks down to the diagonal 
$\mathbb{Z}_2$, which hosts an SWSSB-ASPT order, as observed in the previous model.
The critical point $J=1$ 
separates these two phases and exhibits 
a critical behavior 
in the correlation $\braket{\sigma_{i}^z\sigma_{i+r}^z}$,
along with
a long-range order in the string order correlation 
$\braket{\sigma_{i-1/2}^z \hat{X}_{i}^2\cdots \hat{X}_{i+r}^2\sigma_{i+r+1/2}^z}$. 
We summarize various correlation functions of this model at criticality in Table \ref{tb:correlation}.

\section{Doubled state picture}
We have seen that ground states of $\mathbb{Z}_2$ gauge theories can be interpreted as purified states of the corresponding mixed states. 
Purification is a common technique to treat mixed states as pure states. On the other hand, mixed states can be mapped to pure states in the doubled Hilbert space in a canonical way by the Choi-Jamio\l kowski isomorphism \cite{CHOI1975285,
JAMIOLKOWSKI1972275}.
To see the isomorphism, note that a density matrix is an element of $\operatorname{End}(\mathcal{H})$, and the isomorphism maps the element to $\mathcal{H}\otimes\mathcal{H}^*$. For example, a pure state $\ket{\psi}\bra{\psi}$ is mapped to $\ket{\psi}\otimes\ket{\psi}^*\in\mathcal{H}
\otimes {\cal H}^*$. We denote an operator that acts on $\mathcal{H}$ by using the tilde.
Let $\rho_0$ be a pure state and $\rho=\mathcal{E}_{ZZ}(\rho_0)$.
We denote the doubled state obtained from $\rho$ $(\rho_0)$ by $\kket{\rho}$ $(\kket{\rho_0})$. Since both $\rho_0$ and $\rho$ have the strong $\mathbb{Z}_2$ symmetry, the corresponding states $\kket{\rho_0}$, $\kket{\rho}$ have a $\mathbb{Z}_2\times\mathbb{Z}_2$ symmetry generated by $\prod_{j}\sigma_{j}^x,\prod_{j}\widetilde{\sigma}_{j}^x$.

\begin{figure*}
\includegraphics[scale=0.5]{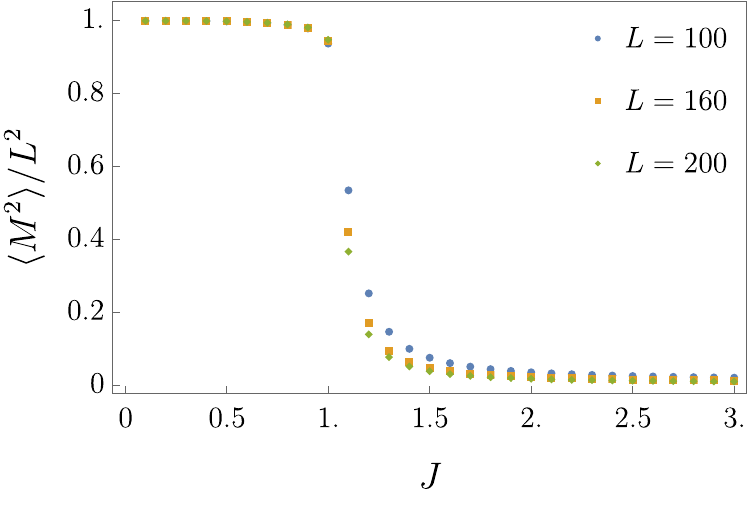}
\includegraphics[scale=0.5]{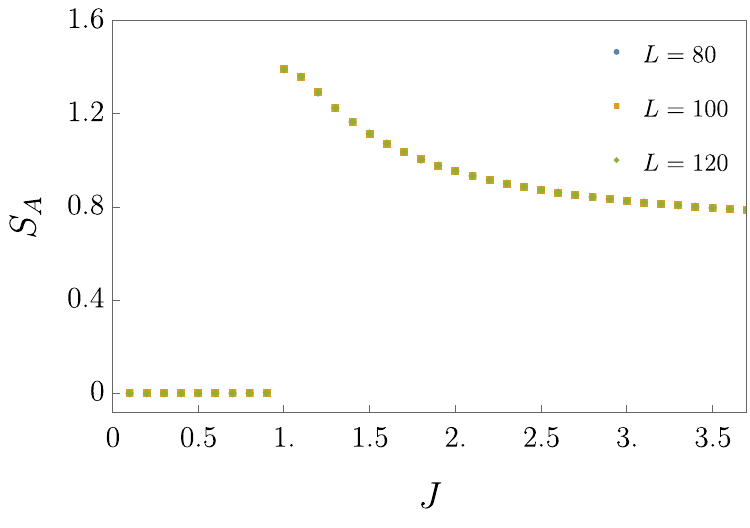}
\caption{
The magnetization squared and entanglement entropy for the mixed state $\varrho$ 
obtained from the model
\eqref{eq:TFI_eff},
$\varrho= \mathrm{Tr}_{{\cal H}_A}(\rho_0)$.
Here, 
the mixed state $\varrho$
can also be obtained from 
the ground state
$\tilde{\rho}_0$
of 
\eqref{model 1 unitary gauge}
and applying the decoherence channel, 
${\cal E}_{ZZ}$,
$\varrho = {\cal E}_{ZZ}(\tilde{\rho}_0)$.
The corresponding state 
$|\varrho\rangle\!\rangle$
in the double state picture 
can be obtained 
numerically 
by the density matrix renormalization group (DMRG)
using iTensor library \cite{itensor, itensor-r0.3}.
The magnetization squared 
can be evaluated as 
$
\langle M^2 \rangle\equiv
\langle\!\langle 
\varrho|
(\sum_j \sigma^z_{j-1/2})^2 
|\varrho\rangle\!\rangle$.
The entanglement entropy 
($S_A$) is calculated from 
$|\varrho\rangle\!\rangle$
by tracing out the half of the chain
in the double Hilbert space. 
Both numerical simulations are implemented with open boundary conditions.
\label{fig1}}
\end{figure*}

How can we understand correlations for mixed states 
in the double state picture? 
Through a simple calculation,
we find that 
\begin{equation}
    \operatorname{Tr}(\rho 
    \sigma^z_{i}
    \sigma^z_{i+r})
    \propto
    \bbra{\rho_0}
    \sigma^z_{i}\sigma^z_{i+r}\kket{\rho}
    =\bbra{\rho_0}
    \widetilde{\sigma}^z_{i}
    \widetilde{\sigma}^z_{i+r}\kket{\rho}.
\end{equation}
On the other hand,
\begin{equation}
    \operatorname{Tr}(\rho 
    \sigma^z_{i}
    \sigma^z_{i+r}\rho 
    \sigma^z_{i}
    \sigma^z_{i+r})\propto \bbra{\rho}
    \sigma^z_{i}
    \widetilde{\sigma}^z_{i}
    \sigma^z_{i+r}
    \widetilde{\sigma}^z_{i+r}\kket{\rho}
\end{equation}
(Table \ref{Table IV}).
Since the operator $\sigma^z_{i}\widetilde{\sigma}^z_{i}$ is 
charged only for the off-diagonal global symmetry, this correlation diagnoses whether the off-diagonal symmetry is spontaneously broken or not. However, as we have already discussed, this R\'enyi-2 correlator is always nontrivial and so the off-diagonal symmetry is necessarily broken in $\kket{\rho}$. 
On the other hand, SSB of the diagonal $\mathbb{Z}_2$ symmetry is diagnosed by  $\bbra{\rho}\sigma^z_{i}\sigma^z_{i+r}\kket{\rho}$. 
However, it is proportional to $\operatorname{Tr}(\rho^2 \sigma^z_{i}\sigma^z_{i+r})$ and the expectation value depends on the specific model in general. We have numerically calculated 
the magnetization and entanglement entropies of Model 1
we discussed above 
(Fig.\ \ref{fig1}).
We found that there is an order one magnetization squared and entanglement entropies obey the area law. 
In particular, the entanglement entropy is about $2\ln(2)$.

\section{Generalizations}
While we have hitherto discussed systems in one spatial dimension, our formulation can be generalized to higher dimensions. Specifically, in a spatial $d$-dimensional system, 
the Gauss law for the vertex $v$ is imposed to be 
$X_{v}\prod_{l}\sigma^z_{l}\ket{\text{phys}}=+\ket{\text{phys}}$
where $l$ denotes the links adjacent to the vertex. In general, such a gauge theory has a magnetic $\mathbb{Z}_2$ 
$(d-1)$-form symmetry and its charged object, which is a $(d-1)$-dimensional object, is composed of the $Z_{l}$ operators. 
This object serves as an order parameter for the $(d-1)$-form symmetry. 
As in the case of $(1+1)d$, 
mixed states obtained by tracing out 
the vertex degrees of freedom 
can be related to a particular quantum channel.
This quantum operation exactly preserves the correlations of the order parameter and completely breaks the strong symmetry spontaneously.
The idea discussed in this paper can readily be generalized to other gauge theories such as for finite gauge groups and higher-form gauge symmetries.

\section{Discussions} 
In this paper, 
we present a method to construct 
various phases with the spontaneous breaking of strong symmetry.
Taking the partial trace is equivalent to applying the quantum operation $\mathcal{E}_{ZZ}$ in the unitary gauge. A key property enabling us to calculate correlation functions is that the ``decohering'' operator $\mathcal{E}_{ZZ}$ commutes with the order parameter of charged operators for global symmetries.
This commutativity is always satisfied when decohering operators are obtained in the aforementioned setup, namely, written as tracing out matter degrees of freedom of lattice gauge theories.
We expect that our construction can be extended to a wider class of lattice models and decoherence channels -- we leave it for future work.

One of the merits of our construction is that 
we can apply various knowledge 
of gauge theories through the correspondence between 
gauge theories and mixed states. 
This can help us to study some aspects of mixed states of matter. 
For example, we expect we can explore the ``duality web'' in mixed states, as the web for (topological) gauge theories is already well-studied. 
In a mixed-state picture, duality operations are replaced by 
appropriate quantum operations as discussed in 
e.g.~\cite{Okada:2024qmk, Khan:2024lyf}.

In this paper, we have characterized the criticality of mixed states by the algebraic decay of correlation functions. On the other hand, recent studies have proposed definitions of gaps in mixed-state phases \cite{Ma:2023rji, Sang:2024vkl}. Understanding the critical states studied in this paper in terms of these various notions of 
gap criteria is an intriguing direction for future work.

{\it Note added}: While the preparation of the draft was at the final stage, we learned \cite{Guo:2024ecx} in which
some of the strong SSB phases 
and criticalities
in this paper 
were also discussed from the perspective of 
the imaginary time evolution of Lindbladians.

\section*{Acknowledgments}
T.A.~thanks Kenji Shimomura for useful discussions. 
We acknowledge the workshop ``Recent Developments and Challenges in Topological Phases'' (YITP-T-24-03) held at Yukawa Institute for Theoretical Physics (YITP), Kyoto University.
T.A.~is supported by JST CREST (Grant No.~JPMJCR19T2) and JSPS KAKENHI Grant Number 25KJ1557.
S.R.~is supported by Simons Investigator Grant from the Simons Foundation (Grant No.~566116).
M.W.~is supported by a Grant-in-Aid for JSPS Fellows (Grant No.~22KJ1777), a Grant-in-Aid for Early-Career Scientists (Grant No.~25K17387) and by MEXT KAKENHI Grant (Grant No.~24H00957).

\appendix



\section{Proof of the main argument}\label{sec:proof_main}

In this appendix, 
we give the proof of \eqref{eq:main_eq}.
In the following,
we will work with the periodic boundary condition and the $z$ basis,  
\begin{equation}
    Z_{j}\ket{a,b}=(-1)^{a_{j}}\ket{a,b},\quad \sigma_{j+1/2}^z\ket{a,b}=b_{j+1/2}\ket{a,b},
\end{equation}
where $a_{j}=0,1$ and $b_{j+1/2}=\pm 1$.
Here, 
$|a\rangle$ is the basis of
${\cal H}_A$,
and $Z_j$ acts faithfully on
${\cal H}_A$. 
For Model 1 in the main text, 
${\cal H}_A={\cal H}_v$.
When $\mathcal{H}_{A}\subsetneq \mathcal{H}_{v}$
(Model 2 and 3), 
$Z_{j}$ can be understood as the tensor product of $Z_{j}$ and the identity operator on 
the orthogonal complement of ${\cal H}_A$.
In such cases, the basis is specified by $\{a_j\}_{j}$ 
along with eigenvalues in the orthogonal complement of ${\cal H}_A$, 
which we omit for simplicity.

We note that $U_{CZ}$ acts on the basis as
\begin{equation}
    U_{CZ}\ket{a,b}=U_{CZ}^\dagger\ket{a,b}=\prod_{j=1}^{L}(b_{j-1/2}b_{j+1/2})^{a_{j}}\ket{a,b}.
\end{equation}
Using the $z$ basis, 
the partial trace of $\rho$ can be expanded as
\begin{equation} \label{aeq:trace}
    \operatorname{Tr}_{\mathcal{H}_{A}}(\rho)=\sum_{a,b,b^\prime}\ket{b}\bra{a,b}\rho\ket{a,b^\prime}\bra{b^\prime}.
\end{equation}
(Here, we fix the index 
of the orthogonal complement 
of ${\cal H}_A$,
which is not shown explicitly.
Or equivalently, one can regard that the index $b$ 
carries this information implicitly.)
On the other hand,
the action of 
$\operatorname{Tr}_{\mathcal{H}_{A}}\left(\mathcal{E}_{CZ}(\rho)\right)$ can be expressed as
\begin{widetext}
    \begin{align}
        \operatorname{Tr}_{\mathcal{H}_{A}}\left(\mathcal{E}_{CZ}(\rho)\right)    &=\sum_{a,b,b^\prime}\ket{b}\bra{a,b}U_{CZ}\,\rho\,U_{CZ}^\dagger\ket{a,b^\prime}\bra{b^\prime}\\
        &=\sum_{a,b,b^\prime}\ket{b}\bra{a,b}\prod_{j}(b_{j-1/2}b_{j+1/2})^{a_{j}}(b_{j-1/2}^\prime b_{j+1/2}^\prime)^{a_{j}}\,\rho\ket{a,b^\prime}\bra{b^\prime}.
    \end{align}
The subsequent action of 
$\mathcal{E}_{ZZ,j^\prime}$
can be calculated as
    \begin{align}
        \begin{split}
        \mathcal{E}_{ZZ,j^\prime}\left(\operatorname{Tr}_{\mathcal{H}_{A}}\left(\mathcal{E}_{CZ}(\rho)\right)\right)
        =& \frac{\operatorname{Tr}_{\mathcal{H}_{A}}\left(\mathcal{E}_{CZ}(\rho)\right)
        +\sigma^z_{j^\prime-1/2}
        \sigma^z_{j^\prime+1/2}\operatorname{Tr}_{\mathcal{H}_{A}}\left(\mathcal{E}_{CZ}(\rho)\right)
        \sigma^z_{j^\prime-1/2}\sigma^z_{j^\prime+1/2}}{2}\\
        =&\sum_{a,b,b^\prime}\ket{b}\bra{a,b}\prod_{j\neq j^\prime}(b_{j-1/2}b_{j+1/2})^{a_{j}}(b_{j-1/2}^\prime b_{j+1/2}^\prime)^{a_{j}}\,\rho\ket{a,b^\prime}\bra{b^\prime}\\
        &\times (b_{j^\prime-1/2}b_{j^\prime+1/2})^{a_{j^\prime}}(b_{j^\prime-1/2}^\prime b_{j^\prime+1/2}^\prime)^{a_{j^\prime}}\frac{1}{2}\left(1+b_{j^\prime-1/2}b_{j^\prime+1/2}b_{j^\prime-1/2}^\prime b_{j^\prime+1/2}^\prime\right).
        \end{split}
    \end{align}
By repeating this calculation, we obtain
    \begin{equation}
    \mathcal{E}_{ZZ}\left(\operatorname{Tr}_{\mathcal{H}_{A}}\left(\mathcal{E}_{CZ}(\rho)\right)\right)
    =
    \sum_{a,b,b^\prime}\prod_{j}\delta\left(b_{j-1/2}b_{j+1/2}b_{j-1/2}^\prime b_{j+1/2}^\prime-1\right)\ket{b}\bra{a,b}\rho\ket{a,b^\prime}\bra{b^\prime},
    \end{equation}
where the delta function is defined as
    \begin{equation}
    \delta\left(b_{j-1/2}b_{j+1/2}b_{j-1/2}^\prime b_{j+1/2}^\prime-1\right)=
    \begin{cases}
        1,\quad b_{j-1/2}b_{j+1/2}b_{j-1/2}^\prime b_{j+1/2}^\prime=1,\\
        0,\quad b_{j-1/2}b_{j+1/2}b_{j-1/2}^\prime b_{j+1/2}^\prime=-1.
    \end{cases}
    \end{equation}
\end{widetext}
However, if a pure state $\ket{\psi}\bra{\psi}$ satisfies the Gauss law, the term $\bra{a,b}\ket{\psi}\bra{\psi}\ket{a,b^\prime}$ gives the delta function contribution because the configuration of the eigenvalues of vertex spins, in the periodic boundary condition, uniquely determines the value $b_{j-1/2}b_{j+1/2}$ for every $j$ as in the decorated domain wall state. Then we have 
\begin{equation}
    \mathcal{E}_{ZZ}\left(\operatorname{Tr}_{\mathcal{H}_{A}}\left(\mathcal{E}_{CZ}(\rho)\right)\right)=\sum_{a,b,b^\prime}\ket{b}\bra{a,b}\rho\ket{a,b^\prime}\bra{b^\prime}.
\end{equation}
This is the same as \eqref{aeq:trace} and completes the proof of \eqref{eq:main_eq}.

\section{Review of (effective) gauging on the lattice}\label{sec:review_gauge}
\subsection{Gauge fixing}

Let us consider the lattice gauge theory model on the 1d lattice,  
\begin{equation}\label{aeq:TFI_gauge}
    H_{\text{TFI}}^{g}[J]
    =-\sum_{j=1}^{L}
    \left(X_{j}
    +JZ_{j} 
    \sigma^x_{j+1/2}
    Z_{j+1}\right).
\end{equation}
We impose the Gauss law constraint,
\begin{align}
\sigma^z_{j-1/2}X_j \sigma^z_{j+1/2}|\psi\rangle=
+ |\psi\rangle
\end{align}
at each site $j$.
This model can be considered as a gauged version of the transverse-field Ising model,
\begin{equation}
    H_{\text{TFI}}[J]=-\sum_{j=1}^{L}
    \left(X_{j}
    +JZ_{j}Z_{j+1}\right)
\end{equation}

We now transform $H^g_{\text{TFI}}$  using $U_{CZ}$. By noting 
\begin{align}
    \begin{split}
        U_{CZ}X_{j} U_{CZ}^\dagger&=
        \sigma^z_{j-1/2}X_{j} 
        \sigma^z_{j+1/2},\\ 
        U_{CZ}\sigma^x_{j+1/2} U_{CZ}^\dagger&=Z_{j} 
        \sigma^x_{j+1/2}Z_{j+1},
    \end{split}
\end{align}
the Hamiltonian \eqref{aeq:TFI_gauge} is transformed as
\begin{equation}
    U_{CZ}H_{\text{TFI}}^{g}[J]\,U_{CZ}^\dagger=-\sum_{j=1}^{L}
    \left(\sigma^z_{j-1/2}
    X_{j} 
    \sigma^z_{j+1/2}+J\sigma^x_{j+1/2}\right)
\end{equation}
with the new Gauss law $X_{j}\ket{\psi}=+\ket{\psi}$. Using this Gauss law, we see that the gauged Hamiltonian is equivalent to 
\begin{equation}
    \widetilde{H}_{\text{TFI}}^{g}=-\sum_{j=1}^{L}\left(
    \sigma^z_{j-1/2}\sigma^z_{j+1/2}+J
    \sigma^x_{j+1/2}\right)
\end{equation}
on which we have no Gauss law constraint. We refer to this gauge choice as the \textit{unitary gauge}.

\subsection{Effective gauging}
Instead of imposing the Gauss law strictly, let us consider the Hamiltonian of the form
\begin{align}\label{aeq:TFI_eff}
    \begin{split}
        H_{\text{TFI}}^{g,K}=-\sum_{j=1}^{L}&\left(X_{j}+JZ_{j}\sigma_{j+1/2}^x Z_{j+1}\right)\\
        &-K\sum_{j=1}^{L}\sigma_{j-1/2}^z X_{j}\sigma_{j+1/2}^z
    \end{split}
\end{align}
for a sufficiently large $K>0$. 
The last term is nothing but the Gauss law operator and commutes with other terms by construction. 
Therefore the ground state of this Hamiltonian also satisfies the Gauss law. 
We refer to the procedure of utilizing 
such a Hamiltonian as \textit{effective} gauging. Since we have no Gauss law constraint anymore in this model, 
the operator $\prod_{j}X_{j}$, which acts trivially on the Hilbert space with strict Gauss law constraint, acts faithfully on the entire Hilbert space and it still generates a global $\mathbb{Z}_2$ symmetry. Thus the global symmetry of the effective gauged Hamiltonian \eqref{aeq:TFI_eff} is $\mathbb{Z}_2\times\mathbb{Z}_2$. With respect to this $\mathbb{Z}_2\times\mathbb{Z}_2$ symmetry, the finite depth unitary circuit $U_{CZ}$ can be understood as an SPT entangler. Therefore, the Hamiltonian \eqref{aeq:TFI_eff} and the Hamiltonian 
\begin{align}\label{aeq:TFI_eff_unitary}
        U_{CZ}H_{\text{TFI}}^{g,K}U_{CZ}^\dagger=-\sum_{j=1}^{L}&\left(\sigma_{j-1/2}^z X_{j}\sigma_{j+1/2}^z+J\sigma_{j+1/2}^x\right)
        \nonumber \\
        &-K\sum_{j=1}^{L}X_{j}
\end{align}
differ by the 
presence of a
nontrivial $\mathbb{Z}_2\times\mathbb{Z}_2$ SPT phase. For example, when $J>1$ the ground state of the effective gauged Hamiltonian \eqref{aeq:TFI_eff} is in the nontrivial SPT phase while \eqref{aeq:TFI_eff_unitary} is in the trivial phase. Notably, the model \eqref{aeq:TFI_eff} shows a gapless symmetry-protected topological order at the critical point $J=1$.

\section{Review of Field-theory perspective}

\subsection{Topological response action of effective gauging}
We provide the topological response action of such an effective gauged model in a general setup. Let $\mathsf{D}$ be a spacetime $(d+1)$-dimensional bosonic theory. Suppose that $\mathsf{D}$ has a finite $\Gamma$ symmetry, which fits into the following central extension of groups:
\begin{equation}
    1 \rightarrow A \rightarrow \Gamma \rightarrow G \rightarrow 1.
\end{equation}
We denote the partition function of $\mathsf{D}$ with the background gauge field by $Z_{\mathsf{D}}[G,A]$. We assume that the abelian group $A$ is non-anomalous. If one gauges the $A$ symmetry, 
the partition function of the gauged theory is given by 
\begin{equation}
    Z_{\mathsf{D}/A}[\hat{A},G]=\#\sum_{a}Z_{\mathsf{D}}[G,a]\,e^{2\pi i\int\chi(a,\hat{A})},
\end{equation}
where $\chi(\cdot,\cdot)$ denotes a pairing of cochains and $\#$ is a numerical factor that depends on the topology of the spacetime manifold. Now consider effective gauging. The precise topological response action of the effective gauged theory is given by the following form \cite{Ando:2024nlk}.
\begin{align}
    \begin{split}
        Z_{\mathsf{D}/A}[\hat{A},G,A]
        &=Z_{\mathsf{D}/A}[\hat{A},G]\,e^{-2\pi i\int\chi(A,\hat{A})}\\
        &=\#\sum_{a}Z_{\mathsf{D}}[G,a]\,e^{2\pi i\int\chi(a-A,\hat{A})}.
    \end{split}
\end{align}
This expression holds not only for gapped theories but also for gapless phases.

\subsection{Partition functions of gSPTs}\label{intro}
Consider a $(1+1)d$ gapless theory \(\mathcal{G}\) with finite \(G\) symmetry. Here we provide the partition function of a gapless SPT model whose low-energy gapless theory is described by \(\mathcal{G}\). 
We assume that the total symmetry of the gSPT theory is \(\Gamma\), which fits into the following central extension of groups:
\begin{equation}\label{aeq:ses_group}
    1\rightarrow A\rightarrow\Gamma\rightarrow G\rightarrow 1.
\end{equation}
We denote the second cohomology class that specifies the sequence by \([e]\in H^2(G,A)\cong H^2(BG,A)\). Then the partition function of the gapless SPT model is given by
\begin{equation}\label{aeq:gSPT_Z}
    Z_{\text{gSPT}}[A,G]=Z_{\mathcal{G}}[G]\,e^{2\pi i\int \chi(A,G)},
\end{equation}
where \(A\) and \(G\) are background gauge fields for \(A\) and \(G\) group symmetry respectively, and $Z_{\mathcal{G}}[G]$ denotes the partition function of the gapless theory $\mathcal{G}$. As a gSPT theory, we assume that the total $\Gamma$ symmetry is non-anomalous \footnote{The definition of gapless SPTs depends on the literature. Here we assume gSPT to be non-anomalous.}. We say that the gSPT \eqref{aeq:gSPT_Z} is an \textit{intrinsically gapless SPT} (igSPT) if $Z_{\mathcal{G}}[G]$ is anomalous.

\subsection{Construction of gSPTs by effective gauging}
Let us consider a bosonic gapless theory $\mathsf{D}$ with the $\Gamma$ symmetry \eqref{aeq:ses_group}, whose partition function is denoted by $Z_{\mathsf{D}}[G,A]$. We assume that the theory is non-anomalous with respect to $\Gamma$ and consider effective gauging of the $A$ symmetry. The gauged theory has a $\hat{A}\cong A$ symmetry as in the usual gauging. Since we are considering effective gauging where the Gauss law is imposed energetically, the effective gauged theory still has the global $A$ symmetry. As discussed in \cite{Ando:2024nlk}, the partition function of the effective gauged theory is given by 
\begin{align}
    \begin{split}
        Z_{\mathsf{D}/A}[\hat{A},G,A]&=Z_{\mathsf{D}/A}[\hat{A},G]\,e^{-2\pi i\chi\int(A,\hat{A})}\\
        &=\#\sum_{a}Z_{\mathsf{D}}[G,a]\,e^{2\pi i\int\chi(a-A,\hat{A})}.
    \end{split}
\end{align}
When the extension \eqref{aeq:ses_group} is nontrivial, the $Z_{\mathsf{D}/A}[\hat{A},G]$ carries an 't Hooft anomaly \cite{Tachikawa:2017gyf}, but the anomaly is canceled by the factor $e^{-2\pi i\int\chi(A,\hat{A})}$. Thus the total theory $Z_{\mathsf{D}/A}[\hat{A},G,A]$ is non-anomalous, and it is a gSPT theory. In particular, it is an igSPT when the extension \eqref{aeq:ses_group} is nontrivial. Note that the total non-anomalous symmetry of this gSPT is $\hat{A}\times\Gamma$ and the low-energy symmetry group is $\hat{A}\times G$, which has an 't Hooft anomaly in general.

\section{Example of gSPT}
\subsection{Non-intrinsic gSPT}
Consider the case when \(G=1,A=\mathbb{Z}_2\). 
The gapless theory \(\mathcal{G}\) is, for example, realized by the Ising CFT.
\subsection{Intrinsic gSPT} Consider the case when \(G=A=\mathbb{Z}_2,\Gamma=\mathbb{Z}_4\). The igSPT partition function is given by
\begin{equation}
    Z_{\text{igSPT}}[\hat{A},G,A]=Z_{\mathsf{D}/A}[G,\hat{A}](-1)^{\int \hat{A}\cup A},
\end{equation}
where $\mathsf{D}$ is a gapless theory with the non-anomalous $\Gamma$ symmetry. The total symmetry fits into the central extension of the form
\begin{equation}
    1 \rightarrow \mathbb{Z}_2^A \rightarrow \mathbb{Z}_2^{\hat{A}}\times\mathbb{Z}_4^{\Gamma} \rightarrow \mathbb{Z}_2^{\hat{A}}\times\mathbb{Z}_2^{G} \rightarrow 1.
\end{equation}
Due to the nontrivial extension, the cocycle condition for \(A\) is modified as \(\delta A=G^2\). By considering a subgroup such that \(\mathbb{Z}_4=\{(0,0),(1,1),(0,2),(1,3)\}\subset \mathbb{Z}_2^{\hat{A}}\times\mathbb{Z}_4^{\Gamma}, \mathbb{Z}_2=(\mathbb{Z}_2^{\hat{A}}\times\mathbb{Z}_2^{G})_{\text{diag.}}\), this theory can be regarded as the igSPT with the symmetry 
\begin{equation}
    1 \rightarrow \mathbb{Z}_2^{A} \rightarrow \mathbb{Z}_4 \rightarrow \mathbb{Z}_2 \rightarrow 1.
\end{equation} 
Note that in the igSPT theory we discuss here, we can always forget the \(G\) symmetry. Then igSPT becomes a not-intrinsically gSPT with respect to \(\hat{A}\times A\) symmetry.

\section{Explicit density matrix expression of SWSSB-ASPT model}
We provide the explicit density matrix expression of the ``SWSSB-ASPT'' phase discussed in the main text.
Let us start from the density matrix of the cluster state, which is defined as
\begin{equation}
    \rho=\prod_{j=1}^{L}\frac{1+Z_{j}\sigma_{j+1/2}^xZ_{j+1}}{2}\prod_{j=1}^{L}\frac{1+\sigma_{j-1/2}^zX_{j}\sigma_{j+1/2}^z}{2}.
\end{equation}
To calculate $\mathcal{E}_{ZZ}(\rho)$, note that 
\begin{widetext}
    \begin{align}
        \begin{split}
            \mathcal{E}_{ZZ,j^\prime}(\rho)
            &=\frac{\rho+\sigma_{j^\prime-1/2}^z\sigma_{j^\prime+1/2}^z\rho\,\sigma_{j^\prime-1/2}^z\sigma_{j^\prime+1/2}^z}{2}\\
            &=\prod_{j=1}^{L}\frac{1+\sigma_{j-1/2}^zX_{j}\sigma_{j+1/2}^z}{2}\left(\prod_{j\neq j^\prime,j^\prime-1}\frac{1+Z_{j}\sigma_{j+1/2}^xZ_{j+1}}{2}\right)\frac{1+Z_{j^\prime-1}\sigma_{j^\prime-1/2}^x\sigma_{j^\prime+1/2}^xZ_{j^\prime+1}}{4}.
        \end{split}
    \end{align}
\end{widetext}
Thus we obtain
\begin{equation}
    \mathcal{E}_{ZZ}(\rho)=\frac{1+\prod_{j=1}^{L}\sigma_{j+1/2}^x}{2^{L}}\prod_{j=1}^{L}\frac{1+\sigma_{j-1/2}^zX_{j}\sigma_{j+1/2}^z}{2}.
\end{equation}
One can see that 
\begin{equation}
    \frac{\operatorname{Tr}\left(\mathcal{E}_{ZZ}(\rho)\sigma_{i-1/2}^zX_{i}\cdots X_{i+r}\sigma_{i+r+1/2}^z\right)}{\operatorname{Tr}\left(\mathcal{E}_{ZZ}(\rho)\right)}=1,
\end{equation}
i.e., the string order parameter $\sigma_{i-1/2}^zX_{i}\cdots X_{i+r}\sigma_{i+r+1/2}^z$ does not vanish, and also see 
\begin{equation}
    \frac{\operatorname{Tr}\left(\mathcal{E}_{ZZ}(\rho)\sigma_{i+1/2}^z\sigma_{i+r+1/2}^z\mathcal{E}_{ZZ}(\rho)\sigma_{i+1/2}^z\sigma_{i+r+1/2}^z\right)}{\operatorname{Tr}\left(\mathcal{E}_{ZZ}(\rho)^2\right)}=1,
\end{equation}
which indicates spontaneous strong-to weak symmetry breaking of the $\mathbb{Z}_2$ symmetry generated by $\prod_{j}\sigma_{j+1/2}^x$.

\section{Lattice models of gapless SPT}
By effective gauging the TFI model, we obtain the Hamiltonian \eqref{aeq:TFI_eff}
\begin{align}
    \begin{split}
        H_{\text{TFI}}^{g,K}=-\sum_{j=1}^{L}&\left(X_{j}+Z_{j}\sigma_{j+1/2}^x Z_{j+1}\right)\\
        &-K\sum_{j=1}^{L}\sigma_{j-1/2}^z X_{j}\sigma_{j+1/2}^z.
    \end{split}
\end{align}
Note that at $J=1$ this model is gapless and the same as the critical point of \eqref{eq:TFI_eff} and \eqref{eq:gSPT_pure}. 
This model was introduced in \cite{Scaffidi:2017ppg} as a lattice model of a gSPT phase. Note that this model has two $\mathbb{Z}_2$ global symmetry generated by $\prod_{j}\sigma_{j+1/2}^x,\prod_{j}X_{j}$. 
We now explain the ground state of this model is two-fold degenerate under the symmetric open boundary condition. Specifically, consider an open chain $\{1,3/2,\ldots,L,L+1/2\}$ and write a ground state of the model by $\ket{\text{GS}}$.
\begin{align}
        \prod_{j=1}^{L}X_{j}\ket{\text{GS}}&=X_{1}\sigma_{3/2}^z\left(\prod_{j=2}^{L}\sigma_{j-1/2}^z X_{j}\sigma_{j+1/2}^z\right)\sigma_{L+1/2}^z\ket{\text{GS}}
        \nonumber  \\
        &=X_{1}\sigma_{3/2}^z\sigma_{L+1/2}^z\ket{\text{GS}}.
\end{align}
In the last line, we used the fact that the third term in the Hamiltonian commutes with the other terms.
Thus the global symmetry acts on the ground state subspace as 
\begin{equation}
    \prod_{j=1}^{L}\sigma_{j+1/2}^x,\quad X_{1}\sigma_{3/2}^z,\quad \sigma_{L+1/2}^z.
\end{equation}
For boundary interactions under the symmetric boundary condition, we can add arbitrary terms that commute with these operators. However, since the minimum representation of the reduced symmetry algebra is two, the ground state must be degenerate and the degeneracy is protected as long as one imposes the symmetry. In addition to the existence of degenerate edge modes, the gSPT model exhibits a nontrivial expectation value of a string order $\sigma_{i-1/2}^zX_{i}\cdots X_{i+r}\sigma_{i+r+1/2}^z$. These two phenomena, degenerate edge modes and a string order parameter, capture the nontrivial gSPT order.

\section{Analysis of the igSPT model}
Here we give the analysis of the model \eqref{eq:igSPT_pure}:
\begin{align}\label{aeq:igSPT_pure}
    \begin{split}
        \widetilde{H_{3}}=-\sum_{j=1}^{L}&\left(\hat{X}_{j}+J\hat{Z}_{j}\sigma_{j+1/2}^x\hat{Z}_{j+1}^\dagger+K_0\,\sigma_{j-1/2}^z \hat{X}_{j}^2\sigma_{j+1/2}^z\right)\\
        &+\mathrm{h.c.}
    \end{split}
\end{align}
Note that this model is obtained by effectively gauging the $\mathbb{Z}_4$ clock model and was discussed in \cite{Su:2024vrk, Ando:2024nlk} as an igSPT model. To study the model, we first rewrite $\hat{X}_{j},\hat{Z}_{j}$ by using other operators. To do this, it is useful to take an explicit expression of $\hat{X}_{j},\hat{Z}_{j}$ as 
\begin{equation}
    \hat{X}_{j}=\begin{pmatrix}
        0&0&0&1\\ 1&0&0&0\\ 0&1&0&0\\ 0&0&1&0
    \end{pmatrix},\quad 
    \hat{Z}_{j}=\begin{pmatrix}
        1&0&0&0\\ 0&i&0&0\\ 0&0&-1&0\\ 0&0&0&-i
    \end{pmatrix}.
\end{equation}
Here we omit the identity elements that act on other than the $j$-th site. Then we regard the local four-dimensional Hilbert space as $\mathbb{C}^2\otimes\mathbb{C}^2$ and denote Pauli-$\alpha$ matrices acting of the former (latter) $\mathbb{C}^2$ by $\tilde{\sigma}_{j}^\alpha\,(\tilde{\tau}_{j}^\alpha)$.
Specifically, $\hat{X}_{j}^2$ is identified as $\tilde{\sigma}_{j}^x\otimes I_{2,j}$, where $I_{2,j}$ is the $2\times2$ identity matrix and $Z_{j}^2$ is identified as $I_{2,j}\otimes\tilde{\tau}_{j}^z$. Using these operators we can rewrite $\hat{X}_{j},\hat{Z}_{j}$ as follows:
\begin{gather}
    \hat{X}_{j}=I_{2,j}\otimes\left(\frac{1}{2}\tilde{\tau}_{j}^x(1+\tilde{\tau}_{j}^z)\right)+\tilde{\sigma}_{j}^x\otimes\left(\frac{1}{2}\tilde{\tau}_{j}^x(1-\tilde{\tau}_{j}^z)\right),\\
    \hat{Z}_{j}=\tilde{\sigma}_{j}^z\otimes\left(\frac{1}{2}(1+\tilde{\tau}_{j}^z)+\frac{i}{2}(1-\tilde{\tau}_{j}^z)\right).
\end{gather}
Then we see the model \eqref{aeq:igSPT_pure} is equivalent to 
\begin{align}
        \widetilde{H_{3}}=-\sum_{j=1}^{L}&\left(\tilde{\tau}_{j}^x+\tilde{\sigma}_{j}^x\tilde{\tau}_{j}^x+J\tilde{\sigma}_{j}^z\tilde{\sigma}_{j+1}^z(1+\tilde{\tau}_{j}^z\tilde{\tau}_{j+1}^z)\sigma_{j+1/2}^x\right.
        \nonumber \\
        &\left.+K_0\,\sigma_{j-1/2}^z\tilde{\sigma}_{j}^x\sigma_{j+1/2}^z\right).
\end{align}
Since the last term is introduced to gauge effectively, it commutes with other terms. To eliminate these gapped degrees of freedom, we implement the unitary transformation as 
\begin{align}
    \begin{split}
        U_{CZ}\widetilde{H_{3}}U_{CZ}^\dagger
    =-\sum_{j=1}^{L}&\left(\tilde{\tau}_{j}^x+\sigma_{j-1/2}^z\tilde{\sigma}_{j}^x\tilde{\tau}_{j}^x\sigma_{j+1/2}^z\right.\\
    &\left.+J(1+\tilde{\tau}_{j}^z\tilde{\tau}_{j+1}^z)\sigma_{j+1/2}^x+K_0\,\tilde{\sigma}_{j}^x\right),
    \end{split}
\end{align}
where $U_{CZ}$ is with respect to $\tilde{\sigma}_{j}^z$ and $\sigma_{j+1/2}^z$. Now we find that the model is equivalently simplified to 
\begin{equation}
    \widetilde{\widetilde{H_{3}}}
    =-\sum_{j=1}^{L}\left(\tilde{\tau}_{j}^x+\sigma_{j-1/2}^z\tilde{\tau}_{j}^x\sigma_{j+1/2}^z+J(1+\tilde{\tau}_{j}^z\tilde{\tau}_{j+1}^z)\sigma_{j+1/2}^x\right).
\end{equation}
This model has a $\mathbb{Z}_2\times\mathbb{Z}_2$ symmetry generated by $\prod_{j}\tilde{\tau}_{j}^x,\prod_{j}\sigma_{j+1/2}^x$ and exactly solvable. In particular, only $J=1$ is gapless and described by the $U(1)_4$ CFT in the IR.

\bibliography{ref.bib}

\begin{thebibliography}{89}
\bibitem [{com()}]{com:effective_gauge}%
  \BibitemOpen
  \href@noop {} {}\bibinfo {note} {When the matter $\mathbb{Z}_2$ symmetry is a non-trivial subgroup of a larger symmetry group, this ``SPT like'' term carries an 't Hooft anomaly.}\BibitemShut {Stop}%
\end{thebibliography}

\end{document}